\documentclass[12pt,a4paper]{article}
\usepackage[dvips]{graphicx}

\newcommand{\be}{\begin{eqnarray} }
\newcommand{\ee}{\end{eqnarray} }
\newcommand{\beq}{\begin{equation} }
\newcommand{\eeq}{\end{equation} }
\newcommand{\plus}{{\uparrow\uparrow}}
\newcommand{\minus}{{\uparrow\downarrow}}
\newcommand{\simleq}{\; \raisebox{-0.4ex}{\tiny$\stackrel
{{\textstyle<}}{\sim}$}\;} 

\begin{document}
\begin{center}
{\Large \bf
NLO QCD procedure of the SIDIS data analysis with respect to 
light  quark
polarized sea
}
\vskip 1.5cm
{ \large A.N.~Sissakian}
\footnote{E-mail address: sisakian@jinr.ru},  
{\large O.Yu.~Shevchenko}
\footnote{E-mail address: shevch@nusun.jinr.ru},
{\large O.N.~Ivanov}
\footnote{E-mail address: ivon@jinr.ru}\\
\vspace{1cm}
{\it Joint Institute for Nuclear Research}\\

\end{center}

        \vskip 10mm
\begin{abstract}
The semi-inclusive deep inelastic scattering (SIDIS) process is considered.
It is proposed a theoretical procedure allowing  the direct extraction from 
the SIDIS data 
of the first moments of the polarized valence distributions  
and of the first moment difference of the light sea quark polarized distributions
in the next to leading (NLO) QCD order.
The validity of the procedure is confirmed by the respective simulations.
\end{abstract}

\begin{center}
        {\large\bf I. Introduction}
\end{center}
The extraction of the polarized quark and gluon densities is one of the main tasks of 
the SIDIS experiments with the polarized beam and target.
Of a special importance for the modern SIDIS experiments are the questions of strange
quark 
and gluon contributions 
to the nucleon spin, and, also the sea quark share as well as the possibility of broken sea scenario. 
Indeed,
it is known \cite{ref1} that the unpolarized sea of light quarks is asymmetric, 
so that the first moments of the unpolarized $\bar u$ and $\bar d$ quark densities do not equal
to each other:
\be
{ \int_0^1 dx\ [\bar{d}(x)-\bar{u}(x)]=0.147\pm 0.039 \neq 0.\nonumber }
\ee
The question arises: does the analogous situation occurs in the polarized case,
i.e.  whether the polarized density $\Delta\bar u$ and its first moment\footnote{
From now on the notation $\Delta_1 q\equiv\int_0^1dx \Delta q$ will be  used 
to distinguish the local in Bjorken $x$ polarized quark densities $\Delta q(x)$ 
and their first moments.}
$\Delta_1 \bar u\equiv\int_0^1dx \Delta \bar u$ are respectively equal to $\Delta\bar d$ and $\Delta_1\bar d$
or not.


In ref. \cite{ref2}  the possibility of broken sea scenario was analyzed,  considering
the results of SIDIS experiments 
on $\Delta q$ with respect to their consistence with the
Bjorken sum rule (BSR) predictions. It was shown \cite{ref2} that 
using the results of \cite{ref3} 
on the valence quark distributions $\Delta_1 q_V$ obtained in the leading (LO) QCD order,
one can immediately estimate
the first moment difference 
of the $u$ and $d$ sea quark polarized distributions:
\be
\label{eold1}
\Delta_1\bar u-\Delta_1\bar d=0.235\pm0.097.  
\ee
At the same time, it was stressed in \cite{ref2} that this is just a speculation, and, to
get the reliable results on $\Delta q$ from the data obtained at the relatively small 
average $Q^2=2.5\,GeV^2$ \cite{ref3},
one should apply NLO QCD analysis. 
The main goal of this paper is to present such a NLO QCD procedure allowing  the {\it direct}
extraction of the quantity $\Delta_1 \bar u - \Delta_1 \bar d$  from the SIDIS data.

It is known that  the description of semi-inclusive
DIS processes
turns out to be much more complicated in comparison with the
inclusive polarized DIS. First, the fragmentation
functions are involved, for which the
information is limited\footnote {For discussion of this subject
see, for example \cite{ref4} and references therein. }.
Second, the extraction of the quark densities in 
NLO QCD order turns out to be rather difficult,
 since 
the  double convolution products are involved. So, to achieve a
reliable description of the SIDIS data it is very desirable, on the one hand,
to exclude from consideration the fragmentation functions,
whenever possible, and, on the other hand, to try to simplify the NLO considerations as much as possible.

It is well known (see, for example, \cite{ref4} and
references therein) that within LO QCD approximation one can completely exclude
the fragmentation functions from the expressions for the
valence quark polarized distributions $\Delta q_V$
through experimentally measured asymmetries. To this end,
instead of the usual virtual photon asymmetry $A^h_{\gamma
N}\equiv A_{1N}^h$ (which is expressed in terms of the directly
measured asymmetry $A^h_{exp}=
(n^h_{\uparrow\downarrow}-n^h_{\uparrow\uparrow})/(n^h_
{\uparrow\downarrow}+n^h_{\uparrow\uparrow})$
as $A^h_{1N}=(P_B P_T f D)^{-1}A^h_{exp}$), one has to measure
so called "difference asymmetry" $A^{h-\bar h}_N$ which is
expressed in terms of the respective counting rates\footnote{
\label{foot6} As usual, one should realize  the quantities $n^h_{\plus(\minus)}$
entering Eq. (\ref{eold2}) not as the pure event densities but as 
the event densities multiplied by the respective luminosities which, in general,
do not cancel out -- see the Appendix.
} as
\be
\label{eold2}
{
A^{h-\bar{h}}_N(x,Q^2;z)=\frac{1}{P_BP_TfD}\frac{(n^{h}_
{\uparrow\downarrow}-n^{\bar{h}}_{\uparrow\downarrow})-(n^{h}_
{\uparrow\uparrow}-n^{\bar{h}}_{\uparrow\uparrow})}{(n^{h}_
{\uparrow\downarrow}-n^{\bar{h}}_{\uparrow\downarrow})+(n^{h}_
{\uparrow\uparrow}-n^{\bar{h}}_{\uparrow\uparrow})},
} \ee where the event densities $n^h_{\uparrow\downarrow
(\uparrow\uparrow)}=dN^h_{\uparrow\downarrow(\uparrow\uparrow)}
/dz$, i.e. $n^h_{\uparrow\downarrow(\uparrow\uparrow)}dz$ are
 the numbers of
events for anti-parallel (parallel) orientations of incoming lepton and
 target nucleon spins
for the hadrons of type h registered in the interval
 $dz$. Quantities $P_B$ and $P_T$, f and D are
the beam and target polarizations, dilution and depolarization
factors, respectively (for details on these quantities see,
 for
 example, \cite{ref5,smctable} and references therein). Then, the LO theoretical
 expressions
for the difference asymmetries look like (see, for example,
COMPASS proposal \cite{ref6}, appendix A)
\begin{eqnarray}
        \label{enew1}
A_{p}^{\pi^+-\pi^-}&=&\frac{4\Delta u_V-\Delta d_V}{4 u_V -d_V};\quad
A_{d}^{\pi^+-\pi^-}=\frac{\Delta u_V+\Delta d_V}{u_V+d_V}; \\
A_{n}^{\pi^+-\pi^-}&=&\frac{4\Delta d_V-\Delta u_V}
{4d_V-u_V}; \quad
A_{p}^{K^+-K^-}=\frac{\Delta u_V}{u_V}; \quad
A_{d}^{K^+-K^-}=A_{d}^{\pi^+-\pi^-}, \nonumber
\end{eqnarray}
i.e., on the one hand, they contain only valence quark
 polarized densities, and, on the other hand, have the remarkable
 property to be free of any fragmentation functions.
 \begin{center}
         {\bf\large
         II. Theoretical basis of the  procedure
         }
 \end{center}
 Let us start NLO consideration with the known \cite{ref4,ref7,ref8,ref9}
theoretical expressions
for the difference asymmetries
\be
\label{eold3}
A_N^{h-\bar{h}}(x,Q^2;z)=\frac{g_1^{N/h}-g_1^{N/\bar{h}}}{\tilde{F}_1^{N/h}-
\tilde{F}_1^{N/\bar{h}}}\quad (N=p,n,d),
\ee
where the semi-inclusive analogs of the structure functions
 $g_1^N$ and $F_1^N$, functions $g_1^{N/h}$ and ${\tilde F}_1^{N/h}$,
 are related to the respective polarized and unpolarized
 semi-inclusive differential cross-sections as follows \cite{ref8}
\be
\label{eold4}
\frac{d^3\sigma^h_{N\uparrow\downarrow}}{dxdydz}-\frac
{d^3\sigma^h_{N\uparrow\uparrow}}{dxdydz}&=&\frac{4\pi\alpha^2}
{Q^2}\ (2-y)\ g_1^{N/h}(x,z,Q^2),\\
\frac{d^3\sigma^h_N}{dxdydz}&=&\frac{2\pi\alpha^2}{Q^2}\
 \frac{1+(1-y)^2}{y}\ 2\tilde{F}_1^{N/h}(x,z,Q^2).\ee
The semi-inclusive structure functions $g_1^{p(n)/h}$
 are given in NLO by

\begin{eqnarray}
        \label{eold5}
g^{p/h}_1&=&\sum_{q,\bar{q}} e_q^2\Delta q[1+\otimes
 \frac{\alpha_s}{2\pi}\delta C_{qq}\otimes]D^h_{q}
\nonumber\\&+&(\sum_{q,\bar{q}} e_q^2\Delta q)\otimes \frac{\alpha_s}
{2\pi}\delta C_{gq}\otimes D^h_g\nonumber\\&+&\Delta g\otimes
 \frac{\alpha_s}{2\pi}\delta C_{qg}\otimes(\sum_{q,\bar{q}} e_q^2
D^h_{q}),
\end{eqnarray}
\be
\label{eold6}
g_1^{n/h}&=&g_1^{p/h}{\Bigl |}_{u\leftrightarrow d},
\ee
where the double convolution product 
is defined as
\be
\label{eold7}
{
[\Delta q\otimes \delta C\otimes D](x,z)\equiv\int\limits_{\cal \quad D}
\int \frac{dx'}{x'}\frac{dz'}{z'} \Delta q\left(\frac{x}{x'}
\right)\delta C(x',z')D\left(\frac{z}{z'}\right)
}.
\ee

The respective expressions for $2\tilde{F}_1^{p(n)/h}$
have the form analogous to Eq. (\ref{eold5}) with the substitution  $\Delta q
\rightarrow q$, $\delta C\rightarrow \tilde{C}$.
 The expressions for the Wilson coefficients $\delta C_{qq(qg,gq)}$
 and $\tilde{C}_{qq(qg,gq)}\equiv C^1_{qq(qg,gq)}+2(1-y)/
 (1+(1-y)^2)C^L_{qq(qg,gq)}$ can be
 found, for example, in \cite{ref8}, Appendix C.
\\
$\quad\quad$It is remarkable that
due to the properties of the fragmentation functions:
\begin{eqnarray}
        \label{eold8}
D_1&\equiv&D_u^{\pi^+}=D_{\bar{u}}^{\pi^-}=D_{\bar{d}}^{\pi^+}
=D_d^{\pi^-},\nonumber\\
D_2&\equiv&D_d^{\pi^+}=D_{\bar{d}}^{\pi^-}=D_u^{\pi^-}
=D_{\bar{u}}^{\pi^+},
\end{eqnarray}
in the differences
$g_1^{p/\pi^+}-g_1^{p/\pi^-}$ and $
\tilde{F}_1^{p/\pi^+}-\tilde{F}_1^{p/\pi^-}$ (and, therefore, in the
 asymmetries $A_p^{\pi^+-\pi^-}$ and $A_d^{\pi^+-\pi^-}$)
 only the contributions containing the Wilson coefficients $\delta
 C_{qq}$ and $\tilde{C}_{qq}$ survive. However, even then the system of
 double integral equations
\begin{eqnarray}
A_p^{\pi^+-\pi^-}(x,Q^2;z)&=&\frac{(4\Delta u_V-\Delta d_V)
[1+\otimes \alpha_s/(2\pi)\delta C_{qq}\otimes](D_1
-D_2)}{(4u_V-d_V)[1+\otimes \alpha_s/(2\pi)C_{qq}
\otimes](D_1-D_2)},\nonumber\\
A_n^{\pi^+-\pi^-}(x,Q^2;z)&=& A_p^{\pi^+-\pi^-}(x,Q^2;z)|_{u_V
\leftrightarrow d_V} \nonumber
\end{eqnarray}
proposed by E. Christova and E. Leader \cite{ref4}, is rather
difficult to solve directly\footnote{So that it seems that the only real possibility to extract polarized distributions
from the data in NLO QCD order 
is to use some proper fit.
At the same time, it is known that a such procedure is rather ambiguous 
since the sea distributions are very sensitive to
the choice of initial
functions for $\Delta q$  parametrization (and, especially, for $\Delta\bar q $ parametrization).}
with respect to the local quantities
$\Delta u_V(x,Q^2)$ and $\Delta d_V(x,Q^2)$. Besides, the
range of integration ${\cal D}$ used in \cite{ref4} has a very
 complicated form, namely:
\be
\frac{x}{x+(1-x)z}\leq x' \leq 1 \ with \ z\leq z'\leq 1, \nonumber
\ee
if $x+(1-x)z\geq 1$, and, additionally,
 range $$x\leq x'\leq x/(x+(1-x))z$$ with $x(1-x')
/(x'(1-x))\leq z' \leq 1$
if $x+(1-x)z\leq 1$. 
 Such enormous complication of the convolution
 integral range occurs if one introduces (to take into account
the target fragmentation contributions\footnote{Then, one
 should also add the target fragmentation contributions
 to the right-hand side of Eq. (\ref{eold5}).} and to exclude the
 cross-section singularity problem at $z_h=0$) a
 new hadron kinematic variable $z=E_h/E_N(1-x)$
 ($\gamma p$ c.m. frame) instead of the usual semi-inclusive
 variable $z_h=(Ph)/(Pq)=(lab. system)\ E_h/E_\gamma$.
 However, both problems compelling us to introduce
 z, instead of $z_h$, can be avoided (see, for example \cite{ref8,ref9})
 if one, just to neglect the target fragmentation, applies
 a proper kinematical cut $Z<z_h\leq 1$, i.e. properly
 restricts the kinematical region covered by the final state
 hadrons\footnote{This is just what was done in the HERMES and
 SMC experiments, where the applied kinematical cut was
 $z_h>Z=0.2$.}. Then, one can safely use, instead of z,
 the usual variable $z_h$, which at once makes the integration
 range ${\cal D}$ in the double convolution product (9) very
 simple: $x\leq x'\leq 1,\ z_h\leq z'\leq 1$.
 Note that in applying the kinematical cut it is much more
 convenient to deal with the total numbers of events
 (multiplied by the respective luminosities -- see footnote \ref{foot6} and the Appendix)
 \be
 \label{eold9}
 N^h_{\uparrow\downarrow(\uparrow\uparrow)}(x,Q^2)
 {\Bigl |}_Z=\int_Z^1 dz_h\ n^h_{\uparrow\downarrow(\uparrow\uparrow)}
 (x,Q^2;z_h)
 \ee
 within the entire interval $Z\leq z_h\leq 1$ and the
 respective integral difference asymmetries
\footnote{Namely the integral spin symmetries
$ A_{1N}^h=\int_Z^1 dz_h\ g_{1N}^h\left/\int_Z^1dz_h\
 \tilde{F}_{1N}^h\right.$
were measured by SMC and HERMES experiments
(see \cite{ref3,ref5,smctable} and also \cite{ref9}).
 }
\begin{eqnarray}
        \label{enew2}  A^{h-\bar{h}}_N(x,Q^2){\Bigl |}_Z&=&
\frac{1}{P_BP_TfD}\frac{(N^{h}_
{\uparrow\downarrow}-N^{\bar{h}}_{\uparrow\downarrow})-(N^{h}_
{\uparrow\uparrow}-N^{\bar{h}}_{\uparrow\uparrow})}{(N^{h}_
{\uparrow\downarrow}-N^{\bar{h}}_{\uparrow\downarrow})+(N^{h}_
{\uparrow\uparrow}-N^{\bar{h}}_{\uparrow\uparrow})}{\Bigl |}_Z
=\\
\label{eold10}
&=&\frac{\int_Z^1dz_h(g_1^{N/h}-g_1^{N/\bar h})}{\int_Z^1 dz_h(\tilde{F}_1^{N/h}
-\tilde{F}_1^{N/\bar{h}})}\quad (N=p,n,d),\end{eqnarray}
than with the local in $z_h$ quantities $n_{\uparrow\downarrow
(\uparrow\uparrow)}(x,Q^2;z_h)$ and $A_N^{h-\bar{h}}(x,Q^2;z_h)$.
 So, the
 expressions for the proton and deutron integral difference
 asymmetries assume the form\footnote{Here one
 uses the equality $g_1^{d/h}\simeq g_1^{p/h}+g_1^{n/h}$
 which is valid up to corrections of order $O(\omega_D)$,
 where $\omega_D=0.05\pm0.01$ is the probability to find
 deutron in the $D$-state.}
\begin{equation}
        \label{eold11}
A_p^{\pi^+-\pi^-}(x,Q^2){\Bigl |}_Z=\frac{(4\Delta u_V-\Delta d_V)
\int_Z^1 dz_h[1+\otimes\frac{\alpha_s}{2\pi}\delta C_{qq}\otimes]
(D_1-D_2)}
{(4u_V-d_V)\int_Z^1 dz_h[1+\otimes\frac{\alpha_s}{2\pi}\tilde{C}_{qq}
\otimes ](D_1-D_2)},
\end{equation}
\be
\label{eold12}
A_d^{\pi^+-\pi^-}(x,Q^2){\Bigl |}_Z=\frac{(\Delta u_V+\Delta d_V)
\int_Z^1 dz_h[1+\otimes\frac{\alpha_s}{2\pi}\delta C_{qq}\otimes]
(D_1-D_2)}
{(u_V+d_V)\int_Z^1 dz_h[1+\otimes\frac{\alpha_s}{2\pi}\tilde{C}_{qq}
\otimes ](D_1-D_2)},
\ee
where the double convolution product reads
\be
\label{eold13}
{\left[ \Delta q\otimes \delta C\otimes
D\right ] (x,z_h)=\int_x^1\frac{dx'}{x'}\int_{z_h}^1\frac{dz'}{z'}\
\Delta q\left(\frac{x}{x'}\right)\delta C(x',z')D\left(
\frac{z_h}{z'}\right) }. \ee

With a such simple convolution region, one can apply the 
well known property of the n-th Melin moments
$M^n(f)\equiv\int_0^1dx\ x^{n-1}f(x)$
to split the convolution product  into a simple
product of the Melin moments of the respective functions:
\be
\label{eold14}
{
M^n[A\otimes B]\equiv \int^1_0 dx x^{n-1}\int^1_x\frac{dy}
{y}A\left(\frac{x}{y}\right)B(y)=M^n(A)M^n(B).
}
\ee
So, applying the first moment to the difference
asymmetries $A_p^{\pi^+-\pi^-}(x,Q^2){\Bigl |}_Z$ and
 $A_d^{\pi^+-\pi^-}(x,Q^2){\Bigl |}_Z$,
 given by (\ref{eold11}), (\ref{eold12}), one gets
 a system of two 
equations for $\Delta_1 u_V\equiv \int_0^1 dx\ \Delta u_V$ and
$\Delta_1 d_V\equiv \int_0^1 dx\ \Delta d_V$:
\be
\label{eold15}
{
(4\Delta_1 u_V-\Delta_1 d_V)(L_1-L_2)={\cal A}_p^{exp},
}
\ee
\be
\label{eold16}
{
(\Delta_1 u_V+\Delta_1 d_V)(L_1-L_2)={\cal A}_d^{exp},
}
\ee
with the solution
\be
\label{eold17}
{
\Delta_1 u_V=\frac{1}{5}\frac{{\cal A}_p^{exp}+{\cal A}_
d^{exp}}{L_1-L_2};\quad  \Delta_1 d_V=\frac{1}{5}\frac{4{
\cal A}_d^{exp}-{\cal A}_p^{exp}}{L_1-L_2}.
}
\ee
Here we introduce the notation
\begin{eqnarray}
        \label{eold18}
{\cal A}_p^{exp}&\equiv& \int_0^1 dx\ A_p^{\pi^+-\pi^-}
{\Bigl |}_Z
(4u_V-d_V)\int_Z^1 dz_h[1+\otimes \frac{\alpha_s}{2\pi}
\tilde{C}_{qq}\otimes]
(D_1-D_2),\nonumber \\{\cal A}_d^{exp}&\equiv& \int_0^1 dx\
A_d^{\pi^+-\pi^-}{\Bigl |}_Z(u_V+d_V)\int_Z^1 dz_h[1+\otimes
\frac{\alpha_s}{2\pi}
\tilde{C}_{qq}\otimes](D_1-D_2),\end{eqnarray}

\begin{eqnarray}
        \label{eold19}
L_1&\equiv& L_u^{\pi^+}=L_{\bar{u}}^{\pi^-}=L_{\bar{d}}^{\pi^+}
=L_d^{\pi^-},
\nonumber\\
L_2&\equiv& L_d^{\pi^+}=L_{\bar{d}}^{\pi^-}=L_u^{\pi^-}
=L_{\bar{u}}^{\pi^+},
\end{eqnarray}
where
\be
\label{eold20}
{
L_q^h\equiv \int_Z^1 dz_h\left[D_q^h(z_h)+
\frac{\alpha_s}{2\pi}\int_{z_h}^1
\frac{dz'}{z'}\ \Delta_1 C(z')D_q^h(\frac{z_h}{z'})\right]
}
\ee
with the coefficient $\Delta_1 C(z)\equiv \int_0^1 dx\ \delta C_{qq}(x,z)$.

Now one may do the last step to get the NLO QCD equation for the 
extraction of the quantity $\Delta_1\bar u - \Delta_1\bar d $
we are interesting in. Namely, on can use the equivalent of BSR 
(see \cite{ref2} and references therein for details) rewritten in terms of
the valence and sea distributions: 
\be
\label{eold21}
{
\Delta_1\bar{u}-\Delta_1\bar{d}=\frac{1}{2}\ \left |\frac{g_A}{g_V}\right |
-\frac{1}{2}\ (\Delta_1 u_V-\Delta_1 d_V).
}
\ee
Using Eqs. (\ref{eold17}--\ref{eold21}) one gets
a simple expression for the quantity
$\Delta_1\bar{u}-\Delta_1\bar{d}\equiv\int_0^1dx\
(\Delta\bar{u}(x,Q^2)-\Delta\bar{d}(x,Q^2))$ in terms of
experimentally measured quantities, that is valid in NLO QCD :
\be
\label{eold22}
{
\Delta_1\bar{u}-\Delta_1\bar{d}=\frac{1}{2}\ \left |\frac{g_A}{g_V}\right |-\frac{2{\cal A}
_p^{exp}-3{\cal A}_d^{exp}}{10(L_1-L_2)}.
}
\ee
It is easy to see that all the quantities present in the
 right-hand side of (25), with the exception of the  two difference
 asymmetries
 $A_p^{\pi^+-\pi^-}{\Bigl |}_Z$ and $A_d^{\pi^+-\pi^-}
 {\Bigl |}_Z$(entering ${
\cal A}_p^{exp}$ and ${\cal A}_d^{exp}$, respectively) can
 be extracted from  unpolarized \footnote{The
 spin-independent fragmentation functions
 D can be
 taken either from independent measurements of $e^+e^-$ -
 annihilation into hadrons \cite{ref10} or from the hadron production in
 unpolarized DIS \cite{ref11}.} semi - inclusive data
 and can, thus, be considered here as a known input.
 So, the only quantities that have to be measured in 
 polarized  semi-inclusive DIS are the difference
 asymmetries $A_p^{\pi^+-\pi^-}{\Bigl |}_Z$ and
 $A_d^{\pi^+-\pi^-}{\Bigl |}_Z$
 which, in turn, are just simple combinations of the
 directly measured counting rates.

 \begin{center}
         {\bf\large
         III. Errors on the difference asymmetries
         }
 \end{center}
To check the validity of the proposed procedure let us perform the respective simulations.
To this end one can use the polarized event generator  PEPSI \cite{PEPSI}.
First of all, let us  clarify the very important issue of the errors on 
the difference asymmetries. At first sight it could seem that the difference
asymmetries suffer from the much larger errors in comparison with the usual asymmetries.
Indeed, the approximate formula for the estimation of the statistical error 
on the difference asymmetries reads (see Eqs. (A.15), (A.16) in the Appendix)
\be
\label{oshibka}
\delta(A^{\pi^+-\pi^-}_{p(n,d)})\sim{\frac{\sqrt{N^{\pi^+}+N^{\pi^-}}}{N^{\pi^+}-N^{\pi^-}}}.
\ee

So, one can see that, contrary to the usual asymmetries, 
the  difference of the total (for both parallel and anti-parallel beam and target polarizations) 
counting rates for $\pi^+$ and $\pi^-$ production,
$N^{\pi^+}\equiv N_{\minus}^{\pi^+} + N_{\plus}^{\pi^+}$ and $N^{\pi^-}\equiv  N_{\minus}^{\pi^-} + N_{\plus}^{\pi^-}$,
occurs
in the denominator of the expression for  $ \delta A_{p(n,d)}^{\pi^+-\pi^-}$,
and it could lead to the large statistical errors on this asymmetry.
Such situation indeed occurs for the neutron target.
However, fortunately, for the proton and deutron targets 
there is an important circumstance which rescues the situation.

The point is that, unlike the neutron target case (see Fig. 1), 
the production of positive pions on the proton
target (see Fig. 2), 
in the widest region in Bjorken $x$\footnote{
\label{foot13}
This occurs everywhere except for the vicinity of $x_B=0$, where the ratio 
$N^{\pi^+}/N^{\pi^-}$ approaches unity owing to the dominant contribution of the sea quarks \cite{EMC}.
However, let us stress that for the statistical error only  the difference
$N^{\pi^+}-N^{\pi-}$ is of importance, and, at the statistics available to HERMES and COMPASS,
$N^{\pi^+}-N^{\pi-}$ is not a small quantity even in the vicinity of the minimal value $x_B=0.003$  accessible to measurement -- see below.
}, 
essentially exceeds
the production of negative pions, whereas 
for the deuteron target the difference in $\pi^+$ and $\pi^-$ production
is not so drastic but is still essential (see Fig. 3). 

\begin{figure}[htb!]
\begin{center}
\fbox{\includegraphics[height=6.7cm,width=10.7cm]{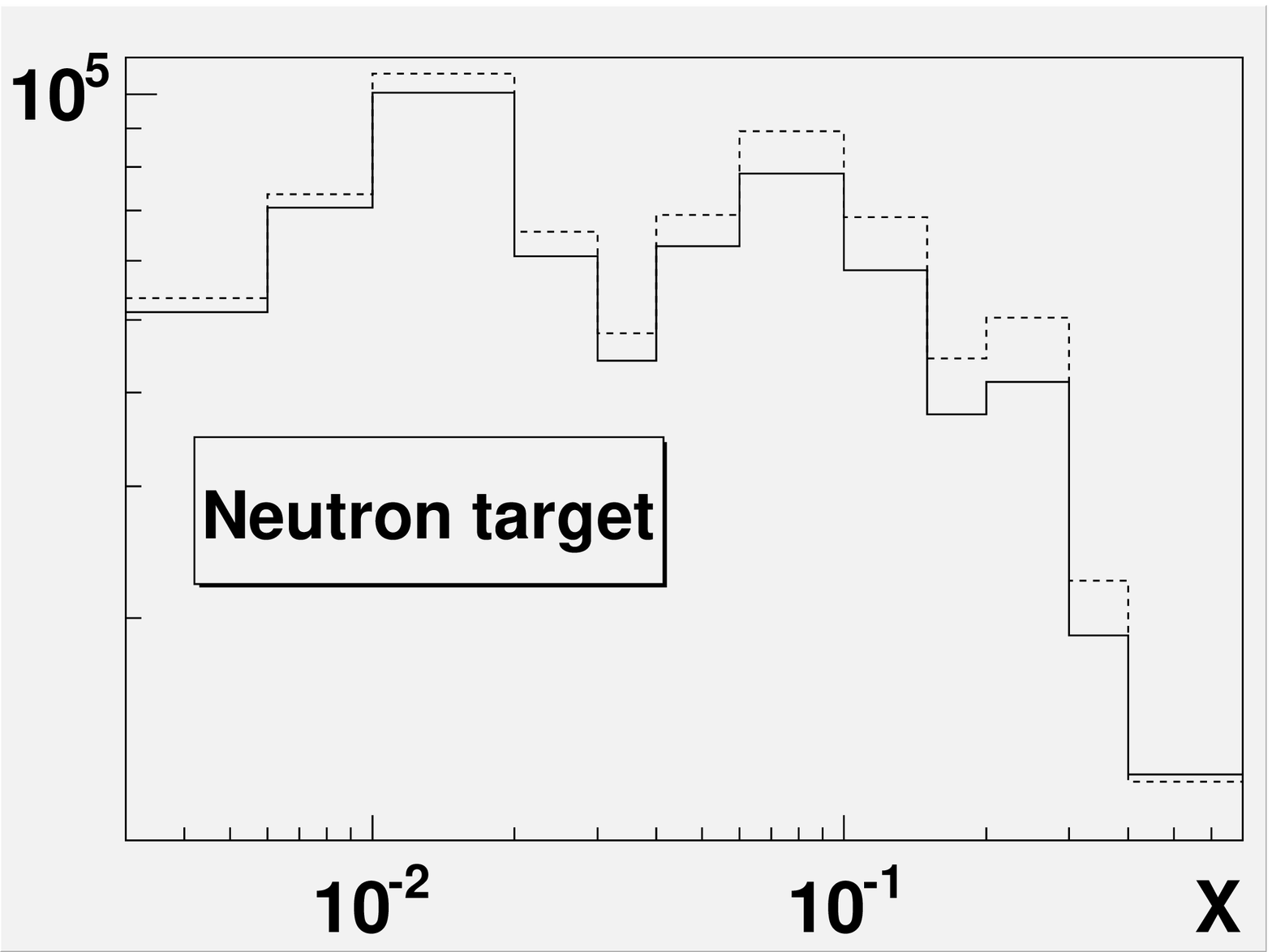}}
\end{center}
\caption{$N^{\pi^+}$ and $N^{\pi^-}$ obtained with PEPSI for neutron target. The dashed and solid lines correspond to $\pi^+$,
and $\pi^-$ productions, respectively.}
\label{fig1}
\end{figure}
\begin{figure}[htb!]
\begin{center}
\fbox{\includegraphics[height=6.7cm,width=10.7cm]{{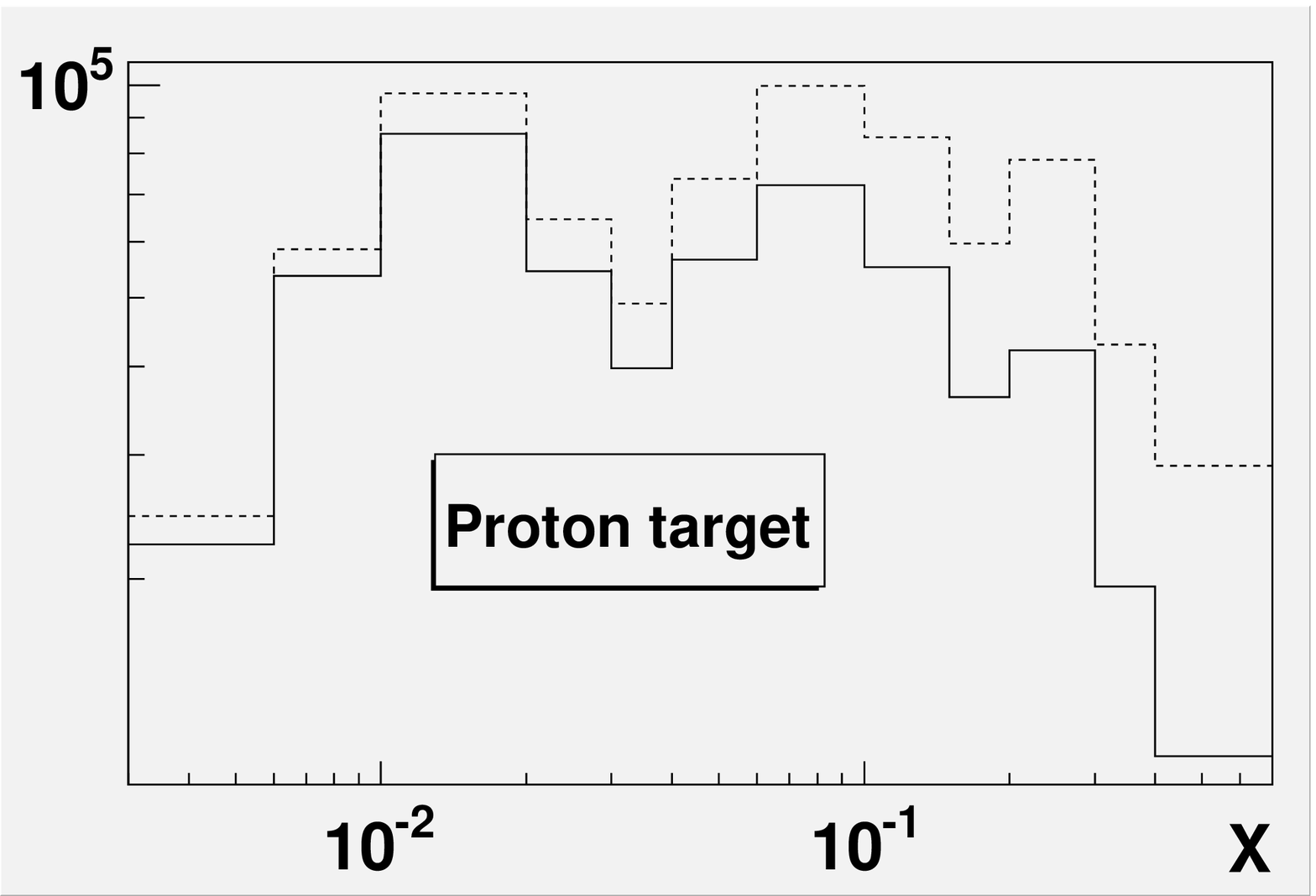}}}
\end{center}
\caption{$N^{\pi^+}$ and $N^{\pi^-}$ obtained with PEPSI  for proton target. The dashed and solid lines  correspond to $\pi^+$,
and $\pi^-$ productions, respectively.
}
\label{fig2}
\end{figure}
\begin{figure}[htb!]
\begin{center}
\fbox{\includegraphics[height=6.7cm,width=10.7cm]{{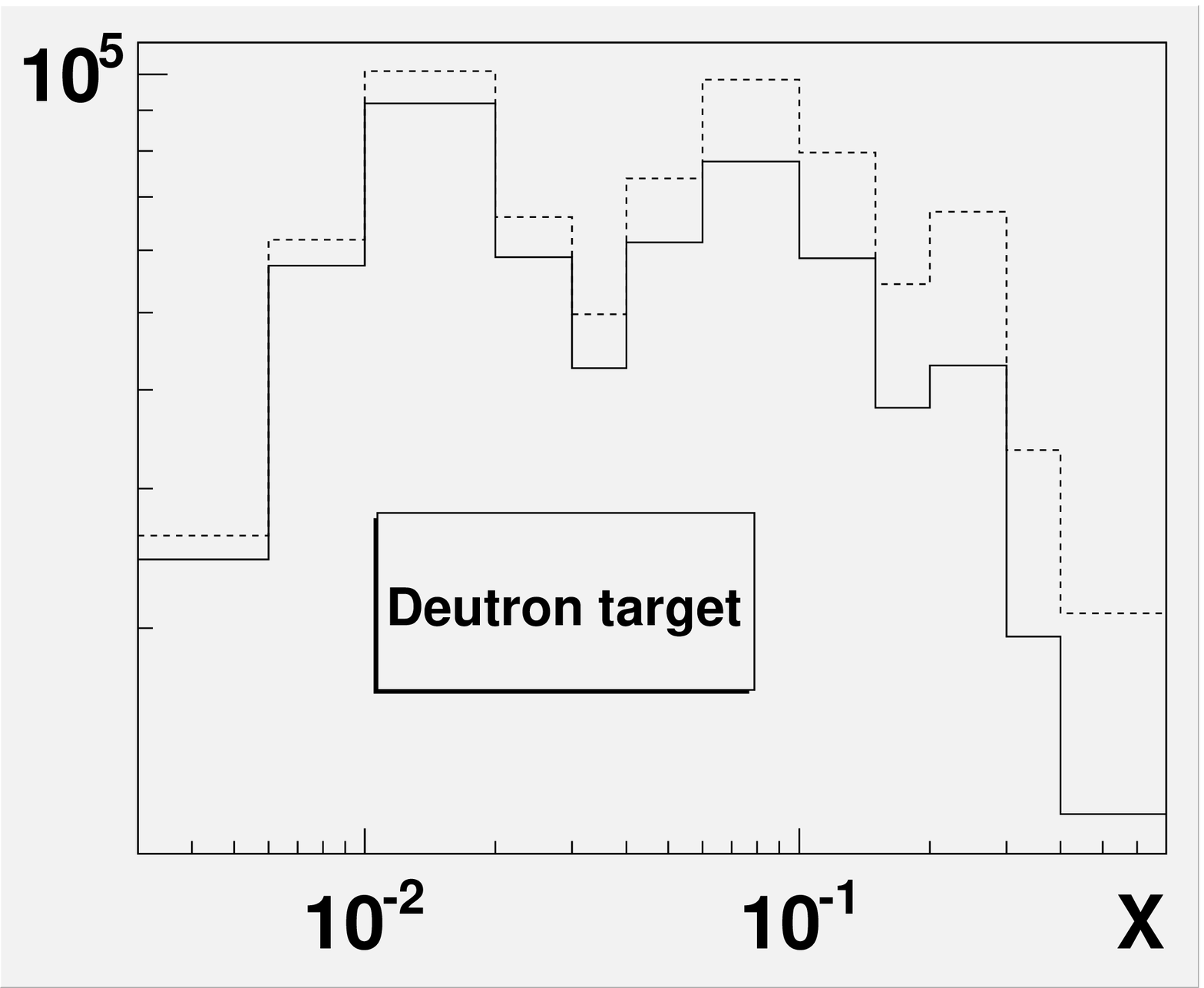}}}
\end{center}
\caption{
$N^{\pi^+}$ and $N^{\pi^-}$ obtained with PEPSI  for deutron target. The dashed and solid lines  correspond to $\pi^+$,
and $\pi^-$ productions, respectively.
}
\label{fig3}
\end{figure}
\begin{figure}[htb!]
\begin{center}
\fbox{\includegraphics[height=6.7cm,width=10.7cm]{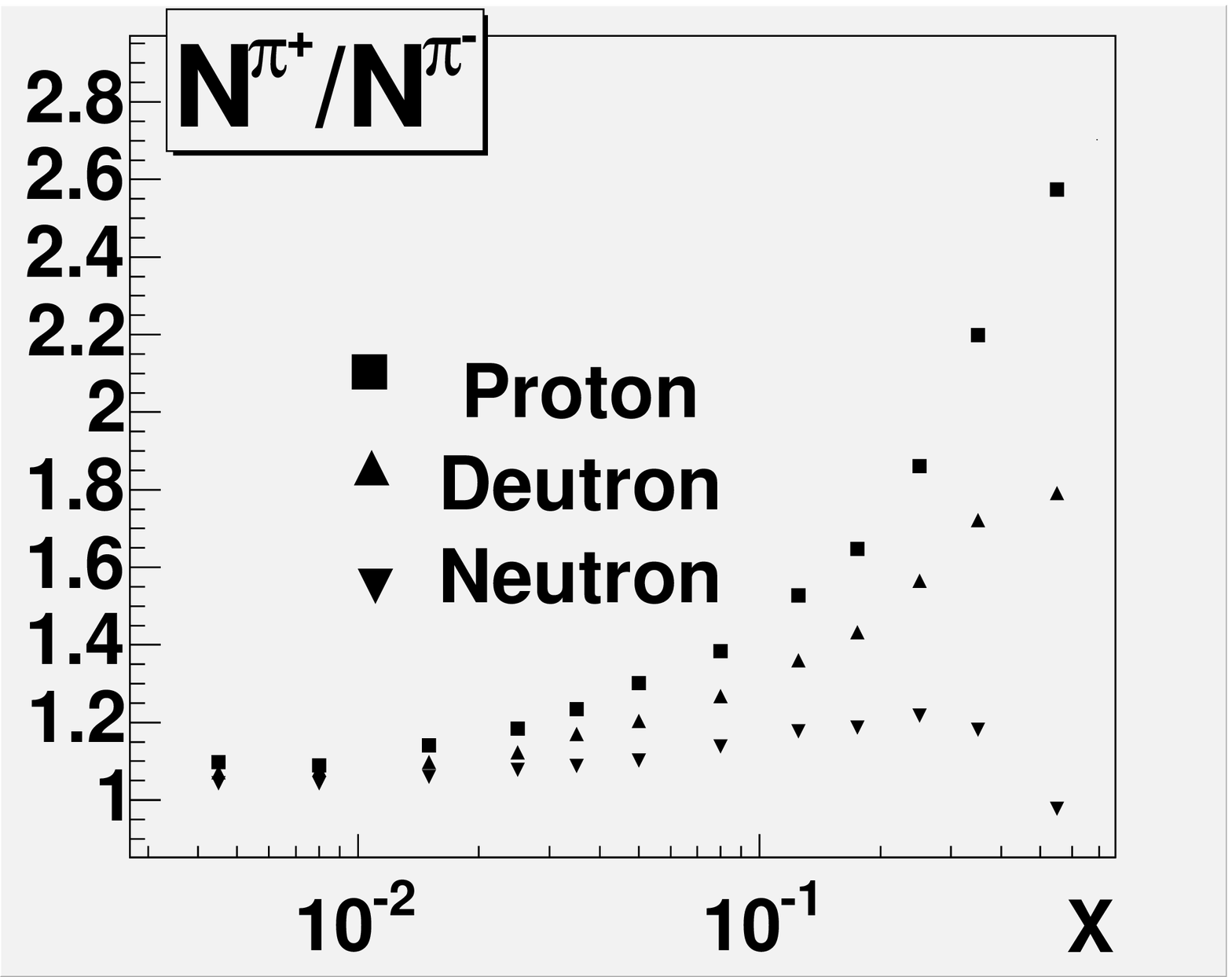}}
\end{center}
\caption {Ratios of 
$N^{\pi^+}\equiv N_{\minus}^{\pi^+} + N_{\plus}^{\pi^+} $ and $N^{\pi^-}\equiv  N_{\minus}^{\pi^-} + N_{\plus}^{\pi^-}$
obtained with the polarized event generator PEPSI
for the different targets. The picture is in good agreement with
the respective EMC result -- Fig. 10 in ref. \cite{EMC}.}
\label{fig4}
\end{figure}

It is of importance that though the histograms in Figs. 1 -- 3 
are obtained using the PEPSI event generator (just summing the events with
parallel and anti-parallel beam and target polarizations), they represent the general, well established experimentally \cite{EMC} 
picture ( see section 4.3 and Fig. 10 in ref. \cite{EMC}), and peculiar to all known SIDIS event generators\footnote{Absolutely
the same histograms for $N_{n,p,d}^{\pi^+}$ and $N_{n,p,d}^{\pi^-}$ are reproduced using, for example, the unpolarized event generator LEPTO \cite{ LEPTO}.}.  
To be sure that concerning the strong  asymmetry between  $N_{p,d}^{\pi^+}$ and  $N_{p,d}^{\pi^-}$
the PEPSI event generator we deal with strictly follows the real \cite{EMC} physical picture, one can also compare
what we have obtained with PEPSI Fig. 4 for the ratio of $N_{p,d}^{\pi^+}$ and $N_{p,d}^{\pi^-}$ with 
Fig. 10 in ref. \cite{EMC}.

Thus, the differences between the total counting rates 
 $N^{\pi^+}\equiv  N_{\uparrow\downarrow}^{\pi^+} + 
N_{\uparrow\uparrow}^{\pi^+}$
and $N^{\pi^-}\equiv  N_{\uparrow\downarrow}^{\pi^-} + 
N_{\uparrow\uparrow}^{\pi^-}$
are not small quantities\footnote{\label{foot15}
Though in the vicinity $x_B=0$ the ratio $N^{\pi^+}/N^{\pi^-}$ approaches unity (see Fig. 4), 
with the applied statistics $3\cdot10^6$ DIS events (absolutely
real for HERMES and COMPASS -- see below),
the difference $N^{\pi^+}-N^{\pi^-}$ entering the error 
significantly differs from zero even near the minimal value
$x_B=0.003$ accessible to measurement. Indeed, in the first bin $0.003<x<0.006$
the PEPSI event generator gives for the proton target $N^{\pi^+}/N^{\pi^-}=1.085$
while $N^{\pi^+}-N^{\pi^-}=2985$, and for deutron target $N^{\pi^+}/N^{\pi^-}=1.053$ while
$N^{\pi^+}-N^{\pi^-}=2020$.
} for both proton and deutron targets,
and, besides, increase with the statistics. As a result, the respective statistical errors turn out to
be quite acceptable.

Let us illustrate this statement by a simple LO example. Using the GRSV2000{\bf LO}(broken sea) \cite{13}
parametrization entering the PEPSI event generator as the input, 
we generate  $3\cdot10^6$ DIS events with $E_{\mu}=160\,GeV^2$ (COMPASS kinematics).
We then construct the "experimental" 
asymmetries  
together with their statistical errors using Eqs. (A.1) and (A.7) from the Appendix, respectively.
These simulated asymmetries are compared with the theoretical ones given by Eqs. (3) -- see Fig. 5.
\begin{figure}[htb!]
\begin{center}
\fbox{\includegraphics[height=10.7cm,width=12.7cm]{{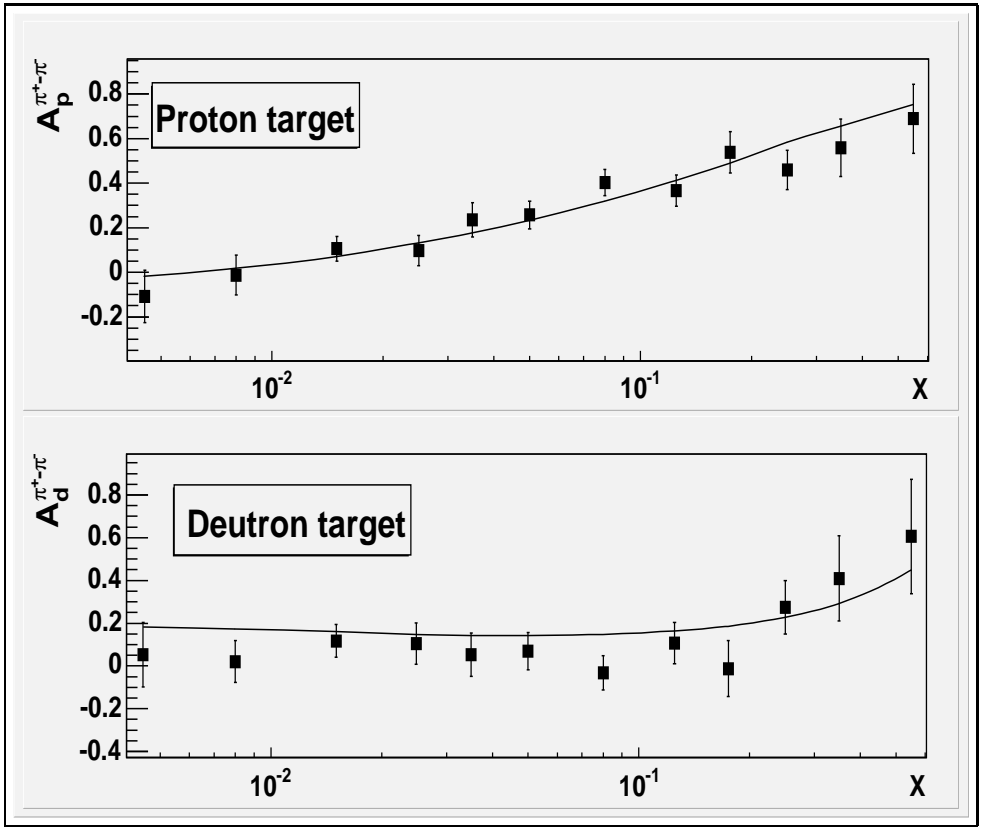}}}
\end{center}
\caption{Simulated (squares) and theoretical  pion difference asymmetries for proton and deutron targets.  
The solid lines correspond to the  
theoretical asymmetries obtained from Eq. (3) with GRSV2000{\bf LO}(broken sea) parametrizations for the valence
distributions.}
\label{fig5}
\end{figure}
One can see that the errors on the simulated asymmetries are quite
acceptable and that the simulated and theoretical asymmetries
are in good agreement within the errors.
Furthermore, 
it is seen from Fig. 6 that the extracted valence\footnote{Namely the valence distributions are essential
for what follows -- see below.} distributions 
are also in good accordance with the respective input parametrizations.
\begin{figure}[htb!]
\begin{center}
\fbox{\includegraphics[height=7.7cm,width=10.7cm]{{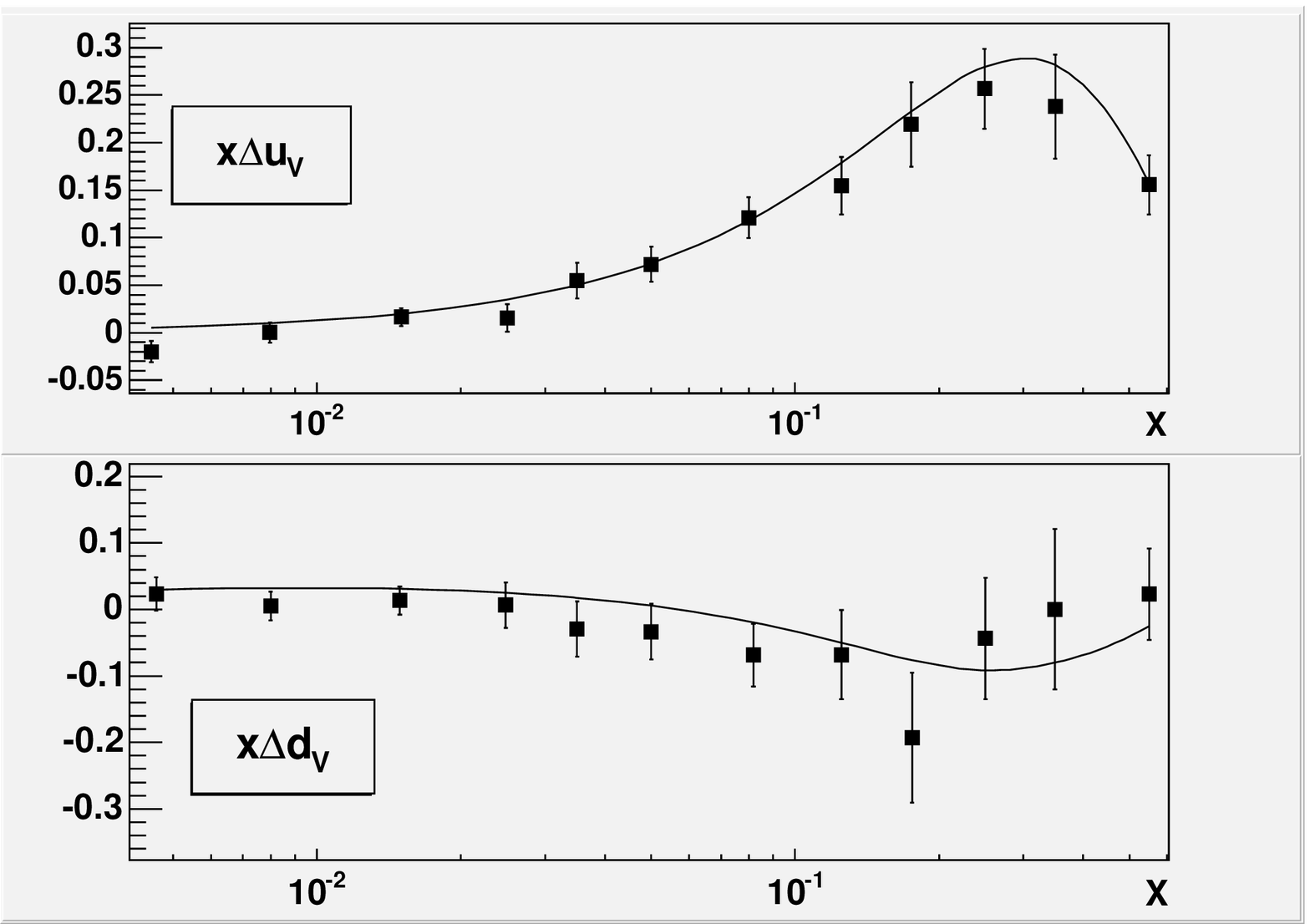}}}
\end{center}
\caption{ The reconstructed polarized valence distributions (squares). The solid lines 
correspond to the respective GRSV2000{\bf LO}(broken sea) parametrizations.}
\label{fig6}
\end{figure}
\begin{center}
        {\bf\large
        IV. Testing of the NLO QCD extraction procedure
        }
\end{center}

\begin{center}
        {\bf
       A. Broken sea scenario 
       }
\end{center}
To perform the NLO QCD analysis, we 
first choose\footnote{\label{foot16}
Note that at present the broken sea scenario is argued as the most probable one \cite{13} 
(see also discussion on this subject at the beginning of this paper and Eq. (\ref{eold1})).}
the GRSV2000{\bf NLO}(broken sea) \cite{13} 
parametrization as an input.
The conditions of simulations
are presented in Table 1 and correspond to HERMES and\footnote{Since the 
COMPASS muon beam energy $160 \,GeV$ is almost the same as SMC one, $190\, GeV$,  
the achieved by COMPASS low $x_B$ boundary at the asymmetry measurements also should be  
about SMC one $0.003$.} 
SMC (COMPASS) kinematics. Let us stress that all the cuts in Table 1 are the standard
cuts applied\footnote{For example, the important cut on invariant mass $W^2>10\,GeV^2$ is applied 
by these collaborations to exclude the
events coming from the resonance region.
} by SMC, HERMES and COMPASS. 
The statistics $3\cdot10^6$ in Table 1 is the total number of DIS events
for both proton and deutron target and for both longitudinal
polarizations. 
Since the statistical error on the pion 
difference asymmetry depends on $N^{\pi^+}$ and $N^{\pi^-}$, one need
to know the respective semi-inclusive statistics -- the total (for all available $x_B$) number  of pions $N_{tot}^{\pi^+}$
and $N_{tot}^{\pi^-}$ corresponding to $1.5\cdot10^6$ DIS events for each target. 
With the all cuts indicated in Table 1,  the PEPSI event generator gives
\begin{eqnarray} 
N_{tot}^{\pi^+}|_{proton}=551281\,,\,\,  N_{tot}^{\pi^-}|_{proton}=358654;\nonumber\quad
N_{tot}^{\pi^+}|_{deutron}=526747\,,\,\, N_{tot}^{\pi^-}|_{deutron}=383826,\nonumber
\end{eqnarray} 
for $E_l=27\,GeV$, while
 \begin{eqnarray} 
 N_{tot}^{\pi^+}|_{proton}=582913\,,\,\,  N_{tot}^{\pi^-}|_{proton}=420709;\nonumber\quad
N_{tot}^{\pi^+}|_{deutron}=559494\,,\,\, N_{tot}^{\pi^-}|_{deutron}=447599,\nonumber
 \end{eqnarray} 
for $E_l=160\,GeV$. It is of importance that these numbers 
are absolutely realistic 
for HERMES (in 2000 HERMES already achieved 
for deutron target $N_{tot}^{\pi^+}=493492$, $N_{tot}^{\pi^-}=402479$ 
-- see Table 5.4 in ref. \cite{GIRL}) and even much less than expected by COMPASS (see COMPASS proposal \cite{ref6}, p. 90 ).

\begin{table}[ht]
        \caption{\footnotesize Simulation conditions. A and B correspond to 
        HERMES and SMC (COMPASS) kinematics, respectively. Here $x_{B}$ and  $x_F$
        are the Bjorken and Feynman $x$ variables, respectively, $z_h=$ is the standard hadronic
        variable and $W$ is the invariant mass of the final hadronic state.
}         
\vskip0.3cm
 \begin{tabular}{|c|c|c|c|c|c|c|}
 \hline
 Kinematics&$E_{lepton}$     & $x_{B}$ &$x_F$&$z_h$&$W^2$&Events\\ 
\hline 
A&27.5 $GeV$ & $0.023<x_{B}<0.6$&$x_F>0.1$&$z_h>Z=0.2$&$W^2>10\,GeV^2$  & $3\cdot10^6$\\ 
\hline 
B&160 $GeV$ & $0.003<x_{B}<0.7$&$x_F>0.1$&$z_h>Z=0.2$&$W^2>10\,GeV^2$ & $3\cdot10^6$\\   

 \hline
 \end{tabular}
\end{table}

To extract the quantities $\Delta_1 u_V$, $\Delta_1 d_V$ 
and, eventually, $\Delta_1 \bar u-\Delta_1 \bar d$ from the simulated asymmetries,
one should first construct the difference asymmetries together with their statistical errors  using Eqs. (A.1) and (A.7) from the Appendix, and then
calculate the quantities ${\cal A}_{p(d)}^{exp}$ and $L_1-L_2$
entering  Eqs. (20), (25). For ${\cal A}_{p}^{exp}$ one should use, instead of integral formula
(\ref{eold18}) the equation (and analogously for ${\cal A}_{d}^{exp}$)
\be
\label{summ}
{\cal A}_p^{exp}&=& \sum_{i=1}^{N_{bins}} \Delta x_i\,  A_p^{\pi^+-\pi^-}(x_i)
{\Bigl |}_Z
(4u_V-d_V)(x_i)\int_Z^1 dz_h[1+\otimes \frac{\alpha_s}{2\pi}
\tilde{C}_{qq}\otimes]
(D_1-D_2),
\ee
where $\Delta x_i$ is the $i-th$ bin width. The parametrizations \cite{Kretzer} 
for the fragmentation functions
and \cite{GRV98} for unpolarized quark distributions are used. Note that here one should not use the 
usual "+"-prescription in the Wilson coefficients $C_{qq}$, 
but its generalization, the so-called "A"-prescription \cite{Ellis}. 
The calculation of $L_1$,$ L_2$ is rather simple and can be done using any numerical method.

Let us introduce the additional notation $\Delta_1^* q_V=\int_{x_{min}}^{x_{max}}dx\,\Delta q_V$
and rewrite\footnote{
As usual,  we neglect contributions to $\Delta_1 q$ from the unmeasured large $x_B$ region
$0.6(0.7)<x_B<1$  because their upper limits given by the unpolarized distributions
are very small there -- see \cite{ref3,smctable}.
} BSR in the form (24) as
\be
\Delta_1\bar u-\Delta_1\bar d=\left[\Delta_1^*\bar u-\Delta^*_1\bar d\right]_{BSR}-\frac{1}{2}\int_0^{x_{min}}dx(\Delta u_V-\Delta d_V),\\
\label{bsrstar}
\left[\Delta_1^*\bar u-\Delta_1^*\bar d\right]_{BSR}=\frac{1}{2}\left|\frac{g_A}{g_V}\right|
-\frac{1}{2}(\Delta_1^*u_V-\Delta_1^*d_V),
\ee
where $x_{min}$ and $x_{max}$ are the boundary points of the available Bjorken $x$
region. It is obvious that dealing with the restricted available
Bjorken $x$ regions, one can directly extract from the measured difference asymmetries 
namely the quantities $\Delta_1^* u_V$, $\Delta_1^* d_V$ and 
$[\Delta_1^* \bar u-\Delta_1^* \bar d]_{BSR}$, while the "tail" contributions 
$\int_0^{x_{min}}dx\,\Delta u_V$, $\int_0^{x_{min}}dx\,\Delta d_V$
and $\frac{1}{2}\int_0^{x_{min}}dx\,(\Delta u_V-\Delta d_V)$ should be studied separately, applying the 
proper extrapolation procedure (see below).

The results on $\Delta_1^* u_V$, $ \Delta_1^* d_V$
and $[\Delta_1^* \bar u-\Delta_1^* \bar d]_{BSR}$
extracted from the simulated difference asymmetries using the presented NLO procedure,
are given in Table 2.

\begin{table}[htb!]
        \caption{\footnotesize GRSV2000{\bf NLO}(broken sea) parametrization. 
Results on $\Delta^*_1 u_V$, $ \Delta^*_1 d_V$
and $[\Delta^*_1 \bar u-\Delta^*_1 \bar d]_{BSR}$
extracted from the simulated difference asymmetries applying the proposed NLO procedure.}
\vskip 0.3cm
 \begin{tabular}{|c|c|c|c|c|}
         \hline
         Kinematics&$Q^2_{mean}$ &$\Delta^*_1 u_V$&$\Delta^*_1 d_V$& $[\Delta^*_1 \bar u-\Delta^*_1\bar d ]_{BSR}$\\
\hline         
A&$2.4\,GeV^2$ &$0.585\pm 0.017$&  $ -0.147\pm0.037 $&$0.268\pm0.020$\\
\hline
B&$7.0\, GeV^2$   &$0.602\pm 0.032$&$-0.110\pm0.080$&$0.278\pm0.040$\\
         \hline

 \end{tabular}
 \end{table}

It is obvious that to be valid,  
the extraction procedure,
being applied to the simulated asymmetries  
should yield results 
maximally close to the ones obtained directly from the parametrization entering the generator as an input.
The  results for the respective parametrization functions integrated over the total $0<x<1$  region in Bjorken 
$x$
and over the regions $0.023< x < 0.6$ (HERMES \cite{ref3} kinematics) and $0.003<x<0.7$ (COMPASS kinematics)
are presented by the Table 3.
\begin{table}[h]
        \caption{\footnotesize Results on $\Delta^*_1 u_V$, $\Delta^*_1 d_V$, $\Delta^*_1 \bar u-\Delta^*_1 \bar d$
        and $[\Delta^*_1 \bar u-\Delta^*_1 \bar d]_{BSR}$ 
obtained from integration of the GRSV2000{\bf NLO}(broken sea) parametrization of the quark distributions
over the total and experimentally available Bjorken $x$ regions. The fifth column is obtained by direct integration
of the respective parametrizations. The sixth column is obtained using BSR and the parametrizations for 
the valence distributions.}  
\vskip 0.3cm
\begin{tabular}{|c|c|c|c|c|c|}
        \hline
        $x_B$&$Q^2$&$\Delta^*_1 u_V$&$\Delta^*_1 d_V$&$\Delta^*_1\bar u-\Delta^*_1\bar d$
&$[\Delta^*_1\bar u-\Delta^*_1\bar d]_{BSR}$\\
        \hline
        $0.0001<x_{Bj}<0.99$&$2.4\,GeV^2$&0.605&-0.031&0.310&0.315\\
        \hline
        $0.023<x_{Bj}<0.6$&$2.4\,GeV^2$&0.569&-0.114&0.170&0.292\\
        \hline
        $0.0001<x_{Bj}<0.99$&$7.0\,GeV^2$&0.604&-0.032&0.309&0.315\\
        \hline
        $0.003<x_{Bj}<0.7$&$7.0\,GeV^2$&0.598&-0.065&0.262&0.302\\
        \hline
       
\end{tabular}
\end{table}

Let us now  compare the results from Tables 2 and 3. 

 First of all notice that
contrary to the actual experiment conditions, the simulations give the possibility 
to check the validity of the extraction method comparing the results of the extraction 
from the simulated asymmetries with an {\it exact} answer. Namely, in our case this is the integral over the 
total region 
of the difference of the parametrizations for $\Delta \bar u$ 
and $\Delta \bar d$ entering the generator as an input:
\be
\label{eold23}
[\Delta_1 \bar u - \Delta_1\bar d]_{exact}\simeq[\Delta^*_1 \bar u - \Delta^*_1\bar d]_{25}
=
\int_{0.0001}^{0.99} dx[\Delta \bar u - \Delta\bar d]_{parametrization}
= 0.310,
\ee
where symbol $[...]_{nm}$ denotes the $n$-th line and $m$-th column of Table 3. 

The second point is that the integral taken {\it directly} (without using BSR) 
from the $\Delta \bar u$ and $\Delta\bar d$  parametrization 
difference over the available to HERMES region 
is almost  two times less than the exact answer (\ref{eold23}):
\be
\label{eold24}
[\Delta^*_1 \bar u - \Delta^*_1\bar d]_{35}
=
\int_{0.023}^{0.6} dx[\Delta \bar u - \Delta\bar d]_{parametrization}
= 0.170.
\ee
This is a direct indication that the HERMES interval in Bjorken $x$ is  too narrow\footnote{ 
Note that the proposed NLO  extraction procedure has nothing to do with that problem -- we just compare
the integrals of the parametrization  over the different Bjorken $x$ regions.} to
extract the quantity $\Delta_1 \bar u - \Delta_1\bar d $ we are interested in directly.

However, there is a possibility to avoid this trouble and essentially
improve the analysis on $\Delta_1\bar u-\Delta_1\bar d$ even with the
narrow HERMES $x_B$ region, applying BSR for $\Delta_1\bar u-\Delta_1\bar d$
extraction.
Indeed, applying Eq. (29) to the HERMES $x_B$ region,
considering that $|g_A/g_V|= 1.2670 \pm 0.0035 $ and calculating the integrals of the valence quark 
parametrizations
over the region $0.023<x<0.6$,
one gets  
\be
\label{eold26}
[\Delta_1^* \bar u - \Delta_1^*\bar d]_{36}^{BSR} = 0.292,
\ee
and this result (contrary to Eq. (\ref{eold24})) is in   good agreement with the exact one, Eq. (\ref{eold23}).

The reason of this good agreement of Eq. (\ref{eold26}) with the exact answer Eq. (\ref{eold23}) is that, 
contrary to the sea distributions, the valence distributions
gather far from the low boundary $x_B=0$ (see, for example \cite{EMC} and references therein).

Thus, this exercise with the integrals of the parametrization functions shows 
that, 
at least within the broken sea scenario, 
the application of Eq. (\ref{eold21}) for $\Delta_1 \bar u - \Delta_1\bar d$ extraction
could give a reliable result on this quantity 
even with the narrow HERMES $x_B$ region. Namely, one should first extract
in the accessible $x_B$ region the truncated moments of the  valence distributions, and  only then
get the quantity $[\Delta^*_1 \bar u - \Delta^*_1\bar d]_{BSR}$ applying Eq. (\ref{bsrstar}).

One can also compare 
elements 55 and 56 from Table 3 corresponding to the SMC (COMPASS)
Bjorken x region with the exact answer, element 45 from the Table 3.
It is seen that 
despite the integral 
over the experimentally available region  taken directly of the sea parametrization difference  
is now much closer to the exact answer, 
 the application of BSR instead of direct extraction 
significantly improves the situation 
even for this much wider 
$x_B$ region.

Returning now to the proposed NLO extraction procedure, let us recall that the application of  BSR  
in the form (\ref{eold21}) (see derivation of Eq. (\ref{eold22}))
is one of the essential elements of the procedure. 
Comparing the result of Table 2 on $[\Delta^*_1\bar u -\Delta^*_1 \bar d]_{BSR}$ obtained from the simulated 
asymmetries with the HERMES kinematics (element 25 from the Table 2)
with both Eqs. (\ref{eold23}) and (\ref{eold26}),
one can see that they are in good agreement with each other.
Besides, comparing the results of Tables 2 and 3 for the COMPASS kinematics, one can see that 
the results on reconstructed $\Delta^*_1 u_V$, $\Delta^*_1 d_V$ and $\Delta^*_1 \bar u-\Delta^*_1 \bar d$
are in still good agreement with the respective
quantities obtained by direct integration of the input parametrization over  both the total $0<x_B<1$
and experimentally available $0.003<x_B<0.7$ regions. 

It is also of importance that even without BSR application, the moments of the valence distributions 
(interesting in themselves) extracted in NLO in the accessible $x_B$ regions, are in a good agreement
with the input parametrization for both HERMES and COMPASS kinematics.

\begin{figure}[htb!]
\begin{center}
\fbox{\includegraphics[height=8.7cm,width=14.7cm]{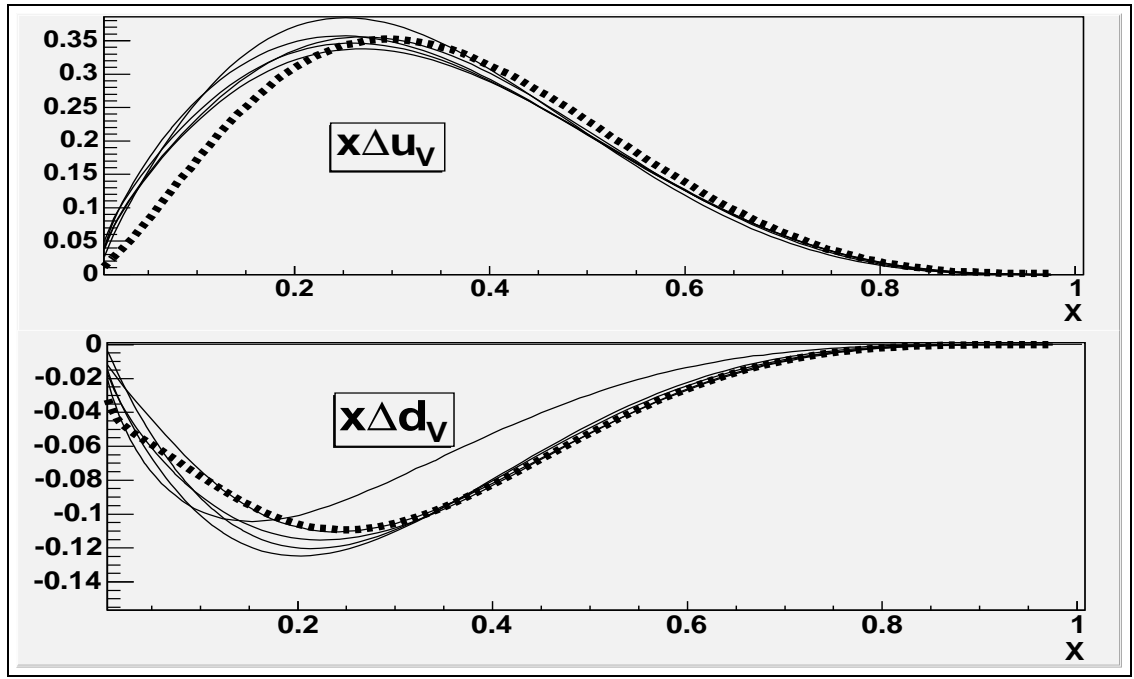}}
\end{center}
\caption{ 
Polarized valence distributions given by the different parametrizations with the symmetric
and weakly broken sea. Solid lines corresponds to 
the purely symmetric parametrizations from refs. \cite{13}
and \cite{PARAMETRIZATIONS}, while the 
dashed line corresponds  to parametrization FS2000 (set i- in ref. \cite{ref9})
with weakly broken sea. All of these parametrizations 
demonstrate quite similar behavior.
}
\label{params}
\end{figure}

\begin{center}
        {\bf
        B. Symmetric sea scenario
        }
\end{center}

Until now we dealt with the broken sea scenario
(which seems to be the most probable one -- see footnote \ref{foot16} ). However, one, certainly, should 
also investigate an alternative opportunity -- the symmetric sea scenario.
Notice that all the known parametrizations
with the symmetric or weakly broken\footnote{The only such parametrization which we know
is FS2000 parametrization \cite{ref9}. $\Delta_1\bar u-\Delta_1\bar d<0.1$ within this parametrization,
and with respect to the valence distribution it behaves quite analogously to the purely
symmetric parametrizations -- See Fig. 7. At the same time the behavior of the parametrization with the strongly
broken sea, GRSV2000 (broken sea),  is absolutely different -- contrary to the parametrizations with the
symmetric or weakly broken sea, $\Delta d_V$ changes the sign at small $x_B$ where the contribution of the sea
quarks become dominant.} polarized sea essentially
differ from the only known at present parametrization with the
strongly broken sea, GRSV2000 (broken sea) we dealt with before.
However,  they rather little differ from each other -- see Fig. 7. 
So, for self-consistence,
we again choose GRSV2000NLO (but with the symmetric sea)
 parametrization as an alternative input. The respective analysis is presented in Table 4.

\begin{table}[h]
        \caption{\footnotesize 
The upper part presents the  results on  
$\Delta^*_1 u_V$, $\Delta^*_1 d_V$ and $[\Delta^*_1 \bar u-\Delta^*_1 \bar d]_{BSR}$ 
 obtained from integration of the GRSV2000{\bf NLO}(symmetric sea) \cite{13} 
parametrization.
The lower part presents the  
results on $\Delta^*_1 u_V$, $ \Delta^*_1 d_V$
and $[\Delta^*_1 \bar u-\Delta^*_1 \bar d]_{BSR}$
extracted from the simulated difference asymmetries applying the proposed NLO procedure
with parametrization GRSV2000{\bf NLO}(symmetric sea)
entering the generator as the input.}

\vskip 0.3cm

\begin{tabular}{|c|c|c|c|c|}
       \hline
         
      $ x_B$ &$Q^2$&$\Delta^*_1 u_V$&$\Delta^*_1 d_V$&$[\Delta^*_1\bar u-\Delta^*_1\bar d]_{BSR}$\\
        \hline
        $0.023<x_{Bj}<0.6$&$2.4\,GeV^2$&0.749&-0.276&0.121\\
        
        \hline
        $0.003<x_{Bj}<0.7$&$7.0\,GeV^2$&0.866&-0.320&0.041\\
      \hline
      $0.0001<x_{Bj}<0.99$&$2.4\,GeV^2$&0.916&-0.339&0.006\\

        \hline
        $0.0001<x_{Bj}<0.99$&$7.0\,GeV^2$&0.914&-0.339&0.007\\

        \hline
        \hline

         Kinematics&$Q^2_{mean}$&$\Delta^*_1 u_V$&$\Delta^*_1 d_V$&$\Delta^*_1 \bar u-\Delta^*_1\bar d$\\
\hline         
A& 2.4&$ 0.736\pm   0.017$& $ -0.310\pm   0.037$&$ 0.111\pm   0.020$\\
     
\hline         
B&$7.0\,GeV^2$&$ 0.842\pm0.032$&$-0.300\pm0.069$&$  0.063 \pm 0.038$\\

         \hline

 \end{tabular}
 \end{table}

Let us analyze the results from Table 4. 
First, one can see that for both A and B kinematics,  the results
of reconstruction in the accessible $x_B$ region
of the all presented in this table quantities are again in good
agreement 
with the input parametrization.  Thus, the analysis performed within
the symmetric sea scenario again confirms that the proposed NLO extraction procedure
satisfies the main criterion of validity -- to reconstruct the quark moments in the experimentally
available $x_B$ region.

On the other hand, performing the reconstruction of the entire 
quantity 
$\Delta_1 \bar u-\Delta_1\bar d$, one, certainly, should not 
roughly put it to $\Delta_1^* \bar u-\Delta_1^*\bar d$. It is necessary to carefully estimate
the unmeasured "tail" 
$\frac{1}{2}\int_0^{x_{min}}dx(\Delta u_V-\Delta d_V)$ entering Eq. (28), especially dealing with so narrow Bjorken $x$
region as the HERMES one. It is clearly seen from the Table 4 where the result on  $\Delta_1^* \bar u-\Delta_1^*\bar d $
is quite close to the exact answer, zero, for the COMPASS $x_B$ region, but in the case of HERMES kinematics
it indicates rather essential deviation from the zero value.

\begin{center}
        {\bf
        C. Low $x_B$ uncertainties
        }
\end{center}

Let us stress that the problem of the unmeasured "tail" estimation is the common and long staying problem
which, however, in any case should be somehow solved if we wish definitely answer the question is the sea symmetric or not.
Nowadays the state of art is such that the polarized SIDIS experiments use the only method of the low $x_B$ contribution
estimations (see, for example, \cite{ref3}, \cite{ref6}): the proper fit to the obtained data on $\Delta q$ is performed
with the subsequent extrapolation of the fitting function
 to unavailable low $x_B$. On the other hand, the low $x_B$ "tails" of all the existing parametrizations 
on $\Delta q$ are obtained using a quite analogous
procedure. Namely, the parametrization on $\Delta q$
is extracted in the 
accessible $x_B$ region  from the fit to the measured inclusive asymmetries and/or structure functions
and then is extrapolated to low $x_B$. It is also of importance that
the degree of the reliability of the low $x_B$ estimations applied in the existing 
parametrizations  increases due to that all the parametrizations
are constructed in the strict accordance with the sum rules on $a_3$ and $a_8$
nonsinglet combinations. Besides, the constructed parametrizations
meet the requirement of agreement with the existing DIS data on $\Gamma_1^{p}$ and $\Gamma_1^{n}$
\cite{LEADER}.

So, we propose to perform the respective estimation
of the quantity $\int_0^{x_{min}}dx(\Delta u_V-\Delta d_V)$  using   the maximal number of the latest available NLO 
parametrizations
. 
The results are presented in the Table 5, 
where the parametrizations from  the refs. \cite{ref9}, \cite{13}  and \cite{PARAMETRIZATIONS} are used.

\begin{table}
        \label{tails}
        \caption{\footnotesize
        Low-$x$ contributions to $\frac{1}{2}(\Delta_1 u_V-\Delta_1 d_V)$ for the different NLO parametrizations.
                }
                \vskip 0.3cm
 \begin{tabular}{|c|c|c|}
 \hline
 NLO parametrization&\multicolumn{2}{|c|}{$\int_0^{x_{min}}dx(\Delta u_V-\Delta d_V)/2$}\\
 \cline{2-3}&$x_{min}=0.023$, $Q^2=2.4\,GeV^2$&$x_{min}=0.003$, $Q^2=7.0\,GeV^2$\\
 \hline
 GRSV2000 (broken sea)&     -0.035 & -0.016\\
 \hline                                                                    
 \hline
 GRSV2000 (symmetric sea)&    0.110 & 0.033\\  
 \hline                                                                    
 FS2000 (i+)   &       0.104       &   0.036     \\
 \hline                                                                    
 FS2000 (i-)   & 0.080             &   0.031     \\
 \hline                                                                    
 LSS2001          &     0.098          &         0.032\\    
 \hline                                                                    
 AAC2000 &     0.116          &         0.046\\
 \hline
 AAC2003 &     0.127          &         0.055\\

 \hline

 \end{tabular}
 \end{table}

Looking at the Table 5 one can conclude that for 
the HERMES $x_B$ region 
\be
\label{bound1}
\frac{1}{2}\int_0^{0.023}dx|(\Delta u_V-\Delta d_V)|\simleq0.13,
\ee
while for the COMPASS $x_B$ region the upper boundary is approximately twice as less:
\be
\label{bound2}
\frac{1}{2}\int_0^{0.003}dx|(\Delta u_V-\Delta d_V)|\simleq0.06.
\ee
Notice that one can estimate only absolute value
of $\int_0^{x_{min}}dx\,(\Delta u_V-\Delta d_V)$,
because we do not know which scenario (symmetric or not) 
is realized in nature. 
For example, the well known  broken sea parametrization GRSV2000
gives the negative sign for this quantity while all the symmetric sea 
parametrizations give the positive sign.
It is also seen
that  the restrictions (\ref{bound1}), (\ref{bound2}) based on the results of Table 5 are rather strong. For safety,
we deliberately  overestimate the upper boundaries choosing the largest numbers from the Table 5
instead of performing the averaging procedure over all used parametrizations (just as it was done
by SMC \cite{smctable}).
Notice also that the restriction (\ref{bound2}) is consistent with the respective
estimation made by SMC (see Table 5 in ref. \cite{smctable}),
whereas the upper boundary given by Eq. (\ref{bound1}) is in four times larger than the HERMES estimation\footnote{
Notice that it is dangerous
to use the simple Regge parametrization for extrapolation in the all rather large $x_B$ region unavailable to HERMES.
The obtained in this way estimations \cite{ref3}  $\int_0^{0.023}dx\,\Delta u_V\simeq0.03$, 
$\int_0^{0.023}dx\,\Delta d_V\simeq-0.03$ seem to be rather underestimated (see discussion on this subject in ref. \cite{ref2}).}.

Thus, to extract the entire quantity $\Delta_1\bar u-\Delta_1\bar d$
we propose to include the upper boundaries on $\frac{1}{2}\int_0^{x_{min}}dx(\Delta u_V-\Delta d_V)$ 
given by  the inequalities (\ref{bound1}) and (\ref{bound2}) into the respective systematical errors,
so that the additional low $x_B$ contributions into the systematical errors of HERMES and COMPASS
look as
\be
\label{deltalow}
\delta_{\,low\, x}{\Bigl |}_{HERMES}=\pm0.13,\\
\label{deltalow1}
\delta_{\,low\, x}{\Bigl |}_{COMPASS}=\pm0.06.
\ee
Certainly, Eqs. (\ref{deltalow}) (\ref{deltalow1}) should not be considered as some strict estimations.
This is just an attempt roughly but with all possible
precautions to estimate could HERMES and (or) COMPASS under their real conditions answer the question is the sea broken
or not.

\begin{center}
        {\bf\large
        V. Discussion and conclusion
        }
\end{center}

With the (rather overestimated) uncertainties given by Eqs. (\ref{deltalow}), (\ref{deltalow1})   
it is quite possible that HERMES would not
see within the total error that $\Delta_1\bar u-\Delta_1\bar d\ne0$, if this quantity
happens too small in reality (for example, about $0.2$). However, if this quantity 
will be about $0.3$ (as it assumed by GRSV2000 broken sea parametrization) and higher,
it could be still possible to see this quantity even with the HERMES $x_B$ region.
On the other hand, it is seen from the Tables 2, 3, 4 and Eq. (\ref{deltalow1})
that the COMPASS $x_B$ region could allow to catch even small difference (if any) between
$\Delta_1\bar u$ and $\Delta_1\bar d$.

In any case, analyzing such a tiny quantity as $\Delta_1\bar u-\Delta_1\bar d $ it is very
desirable to perform a combined analysis with both HERMES and COMPASS data.
For example, having at one's disposal data on the difference asymmetries in the accessible to HERMES 
$x_B$ region, one could involve in the analysis on $\Delta_1\bar u-\Delta_1\bar d$ the respective COMPASS
 data from
the region $0.003<x_B<0.023$.
The point is that a high statistics, especially at low $x_B$ is claimed as one of the COMPASS advantages \cite{ref6} 
as compared with SMC and HERMES experiments.

\vskip 0.3cm
Thus, 
we have tested the proposed NLO QCD extraction procedure 
performing the simulations corresponding to both the broken and symmetric sea 
scenarios.
This analysis confirms that the procedure meets
the main requirement: to reconstruct the quark moments in the 
accessible to measurement $x_B$ region. On the other hand,
even with the overestimated low $x_B$ uncertainty (\ref{deltalow}),
one can conclude that the question 
is $\Delta_1\bar u-\Delta_1\bar d$ equal to zero or not could  be answered even with the
HERMES kinematics in the case of strongly asymmetric polarized sea. 
In any case, the situation is much better with the available to COMPASS $x_B$  region.
\vskip 0.3cm 
\begin{center}
        {\bf\large Acknowledgments}
 \end{center}
  The authors are grateful to R.~Bertini, M.~P.~Bussa,
 O.~Denisov, O.~Gorchakov, A.~Efremov, 
 N.~Kochelev, A.~Korzenev, A.~Kotzinian, V.~Krivokhizhin, E.~Kuraev,
 A.~Maggiora,  A.~Nagaytsev, A.~Olshevsky, 
G.~Piragino, G.~Pontecorvo, J.~Pretz, I.~Savin and  O.~Teryaev,
 for fruitful discussions.
 \begin{center}
         {\bf\large Appendix}
 \end{center}
\renewcommand{\theequation}{A.\arabic{equation}}
  \setcounter{equation}{0}  
Calculating the asymmetries
given by Eq. (\ref{enew2}) together with their statistical errors,
one should have in mind that, contrary to SMC and COMPASS experiments,
in the HERMES  conditions the quantities
$N^{\pi^+(\pi^-)}_{\plus(\minus)}$ entering Eq. (\ref{enew2}) 
are not the pure counting rates, but the counting rates multiplied\footnote{The cancellation
of the luminosities is possible only with the special target setup, like the SMC and COMPASS ones \cite{smctable,ref6}.} 
by the respective luminosities \cite{ref3}. Thus, in general case, 
one should use instead of Eq. (\ref{enew2}) the equation\footnote{Here the beam and target are
assumed to be ideal which means that $P_B=P_T=f=1$. Namely this assumption is adopted in the PEPSI event
generator \cite{PEPSI} we use for simulations.}:
\be
\label{zar}
A_{p(n,d)}^{\pi^+-\pi^-}=\frac{1}{D}\left[\frac{(N_\minus^{\pi^+}-N_\minus^{\pi-})L_\plus-(N_\plus^{\pi^+}-N_\plus^{\pi^-})L_\minus}
{
 (N_\minus^{\pi^+}-N_\minus^{\pi-})L_\plus
+(N_\plus^{\pi^+} -
N_\plus^{\pi^-})L_\minus 
}\right],
\ee
where luminosities $L_{\plus(\minus)}$ are defined as
\be
\label{lumi1}
L_{\plus(\minus)}=(n\Phi)_{\plus(\minus)},
\ee
where $n$ is th area density of the nucleons in the target and $\Phi$
is the beam flux.
Within the paper we do not study the specific peculiarities
of the different experimental setups  and deal only with the event generator where the acceptance $a$ is equal to unity,
so that 
\be
N_{\minus(\plus)}=(n\Phi)_{\minus(\plus)}a\sigma_{\minus(\plus)}\rightarrow(n\Phi)_{\minus(\plus)}\sigma_{\minus(\plus)}
\ee
Thus, Eq. (\ref{lumi1}) for luminosities  is rewritten in the following,
suitable for simulations, form
\be
L_{\plus(\minus)}=N_{\plus(\minus)}/(\sigma_{\plus(\minus)}),
\ee
where $N_{\plus(\minus)}$ are the numbers of inclusive events and $\sigma_{\plus(\minus)}$ are the inclusive
cross-sections automatically calculated by PEPSI (see ref. \cite{PEPSI}) for given sets
of the kinematic conditions.

Choosing as the variables in Eq. (\ref{zar}) the set $N^{\pi^+}_\minus,N^{\pi^-}_\minus,N^{\pi^+}_\plus,N^{\pi^-}_\plus$
and using the general formula (see, for example, \cite{Pretz}) for the statistical error on the function $F$ of variables $x_1,x_2,\ldots$
\be
\label{funcerr}
\delta^2(F(x_1,x_2,\ldots))=\left(\frac{\partial F}{x_1}\right)^2\delta^2(x_1)+\left(\frac{\partial F}{x_2}\right)^2\delta^2(x_2)
+2\left(\frac{\partial F}{x_1}\right)\left(\frac{\partial F}{x_2}\right){\rm cov}(x_1,x_2)+\ldots
\ee
one gets
\be
\label{a6}
\delta^2 (A_{p(n,d)}^{\pi^+-\pi^-})=\frac{1}{D^2Y^4}\left\{(Y-X)^2L^2_\plus[\delta^2 (N^{\pi^+}_\minus)+\delta^2 (N^{\pi^-}_\minus)
-2\,{\rm cov} (N^{\pi^+}_\minus,N^{\pi^-}_\minus)]\right.
\nonumber\\
\left. +(Y+X)^2L^2_\minus[(\delta^2 N^{\pi^+}_\plus)+\delta^2 (N^{\pi^-}_\plus)]-2\,{\rm cov}(N^{\pi^+}_\plus,N^{\pi^-}_\plus)\right\},
\ee
where 
$X$ and $Y$ are the numerator and denominator in the square brackets in Eq. (\ref{zar})
. If the distributions of hadrons $N^{\pi^+}_\minus,N^{\pi^-}_\minus,N^{\pi^+}_\plus,N^{\pi^-}_\plus$ are Poissonian (low multiplicities
$n^{+}$, $n^{-}$ -- see  \cite{Pretz}): 
$\delta(N^{\pi^+(\pi^-)}_{\plus(\minus)})=\sqrt{N^{\pi^+(\pi^-)}_{\plus(\minus)}}$, then one can neglect in Eq. (\ref{a6})  the covariations
${\rm cov}(N^{\pi^+}_\plus,N^{\pi^-}_\plus)$ and ${\rm cov}(N^{\pi^+}_\minus,N^{\pi^-}_\minus)$ with a result: 
\be
\label{a7}
\delta^2 (A_{p(n,d)}^{\pi^+-\pi^-})=\frac{4L_\plus^2L_\minus^2}{D^2}\frac{(N^{\pi^+}_\plus-N^{\pi^-}_\plus)^2[N^{\pi^+}_\minus+ N^{\pi^-}_\minus]
 +(N^{\pi^+}_\minus-N^{\pi^-}_\minus)^2[ N^{\pi^+}_\plus+N^{\pi^-}_\plus]}
 {[(N^{\pi^+}_\minus L_\plus+N^{\pi^+}_\plus L_\minus)-(N^{\pi^-}_\minus L_\plus+N^{\pi^-}_\plus L_\minus)]^4}.
\ee
Operating absolutely analogously, one gets for the error on the usual spin asymmetry the equation
\be
\label{a8}
\delta^2(A_{p(n,d)}^{\pi^+})=\frac{1}{D^2}\frac{4L^2_\minus L^2_\plus(N^{\pi^+}_\minus+N^{\pi^+}_\plus)N^{\pi^+}_\minus N^{\pi^+}_\plus}
{(N^{\pi^+}_\minus L_\plus+N^{\pi^+}_\plus L_\minus)^4}.
\ee
Notice that namely the equation (\ref{a8}) for the  statistical error on the usual semi-inclusive asymmetry 
was used by HERMES \cite{GIRL}.

For the COMPASS experiment $L_\plus=L_\minus$ and
Eq. (\ref{a7}) reduces to
\be
\delta^2 (A_{p(n,d)}^{\pi^+-\pi^-})=\frac{4}{D^2}\frac{(N^{\pi^+}_\plus-N^{\pi^-}_\plus)^2[N^{\pi^+}_\minus+ N^{\pi^-}_\minus]
 +(N^{\pi^+}_\minus-N^{\pi^-}_\minus)^2[ N^{\pi^+}_\plus+N^{\pi^-}_\plus]}
 {[(N^{\pi^+}_\minus +N^{\pi^+}_\plus )-(N^{\pi^-}_\minus +N^{\pi^-}_\plus )]^4}.
\ee
Let us now choose as the variables the set $X_1=N^{\pi^+}_\minus-N^{\pi^-}_\minus$, $X_2=N^{\pi^+}_\plus-N^{\pi^-}_\plus$,
so that the equation (\ref{zar}) for asymmetry is rewritten as
\be
A_{p(n,d)}^{\pi^+-\pi^-}=\frac{1}{D}\frac{X_1L_{\plus}-X_2L_{\minus}}{X_1L_{\plus}+X_2L_{\minus}}.
\ee
Then Eq. (\ref{funcerr}) gives:
\be
&&\delta^2(A_{p(n,d)}^{\pi^+-\pi^-})=\nonumber\\
\label{a11}&&=\frac{4L_{\plus}^2L_{\minus}^2}{D^2}\frac{X_2^2\delta^2(X_1)+X_1^2\delta^2(X_2)}{(X_1L_{\plus}+X_2L_{\minus})^4},
\ee
with 
\be
\label{a12}
&&\delta^2(X_1)=\delta^2(N^{\pi^+}_\minus)+\delta^2(N^{\pi^-}_\minus)-2\,{\rm cov}(N^{\pi^+}_\minus,N^{\pi^-}_\minus),\\
\label{a13}
&&\delta^2(X_2)=\delta^2(N^{\pi^+}_\plus)+\delta^2(N^{\pi^-}_\plus)-2\,{\rm cov}(N^{\pi^+}_\plus,N^{\pi^-}_\plus).
\ee
Again, with a standard assumption that the distributions of hadrons $N^{\pi^+(\pi^-)}_{\plus(\minus)}$ are 
Poissonian: $\delta(N^{\pi^+(\pi^-)}_{\plus(\minus)})=\sqrt{N^{\pi^+(\pi^-)}_{\plus(\minus)}}$, 
one can neglect in Eqs. (\ref{a12}), (\ref{a13})  the covariations
${\rm cov}(N^{\pi^+}_\plus,N^{\pi^-}_\plus)$ and ${\rm cov}(N^{\pi^+}_\minus,N^{\pi^-}_\minus)$.
Then Eq. (\ref{a11}) exactly transforms to Eq. (\ref{a7}).

Let us involve an additional approximation (see \cite{Pretz}, p.7)
\be
\label{virt}
X_1\simeq X_2\simeq \tilde Y/2, 
\ee
where the quantity\footnote{It is easy to see that $\tilde Y$ up to luminosities
coincides with denominator $Y$ in the square brackets in Eq.(\ref{zar}). } $\tilde Y$ is defined as $\tilde Y\equiv(N^{\pi^+}_\minus+N^{\pi^+}_\plus)-(N^{\pi^-}_\minus+N^{\pi^-}_\plus)$
$\equiv N^{\pi^+}-N^{\pi^-}$ .  
Then Eq. (\ref{a11}) reads:
\be
&&\delta^2(A_{p(n,d)}^{\pi^+-\pi^-})=\frac{1}{D^2}\frac{16L_{\minus}^2L_{\plus}^2}{{\tilde Y}^2(L_{\minus}+L_{\plus})^4}\delta^2(\tilde Y)\equiv
\frac{16L_{\minus}^2L_{\plus}^2}{D^2(L_{\minus}+L_{\plus})^4}\frac{1}{(N^{\pi+}-N^{\pi-})^2}\delta^2(N^{\pi+}-N^{\pi-})\nonumber\\
&&=\frac{N^{\pi^+}+N^{\pi^-}}{(N^{\pi^+}-N^{\pi^-})^2}\frac{16L_{\minus}^2L_{\plus}^2}{D^2(L_{\minus}+L_{\plus})^4}.
\ee
With the SMC (COMPASS) target setup $L_{\minus}=L_{\plus}$ and
\be
\label{a16}
\delta^2(A_{p(n,d)}^{\pi^+-\pi^-})=\frac{1}{D^2}\frac{N^{\pi^+}+N^{\pi^-}}{(N^{\pi^+}-N^{\pi^-})^2}.
\ee

It is instructive to reproduce Eq. (\ref{a16}) choosing another variables in Eq. (A1): $\Delta N^{\pi^+}=N^{\pi^+}_\minus-N^{\pi^+}_\plus$,
$\Delta N^{\pi^-}=N^{\pi^-}_\minus-N^{\pi^-}_\plus$, $N^{\pi^+}=N^{\pi^+}_\minus+N^{\pi^+}_\plus$,
$N^{\pi^-}=N^{\pi^-}_\minus+N^{\pi^-}_\plus$, so that with $L_\minus=L_\plus$ (COMPASS target setup) Eq. (\ref{zar})
for asymmetry reads
\be
A_{p(n,d)}^{\pi^+-\pi^-}=\frac{1}{D}\frac{\Delta N^{\pi^+}-\Delta N^{\pi^-}}{N^{\pi^+}-N^{\pi^-}}\equiv\frac{1}{D}\frac{\tilde X}{\tilde Y}.
\ee
Then the respective statistical error look as
\be
&&\delta^2 (A_{p(n,d)}^{\pi^+-\pi^-})=\frac{1}{D^2\,\tilde{Y}^2}\left\{\delta^2(\Delta N^{\pi^+})+\delta^2(\Delta N^{\pi^-})
+\frac{\tilde{X}^2}{\tilde{Y}^2}(\delta^2(N^{\pi^+})+\delta^2(N^{\pi^-}))\right\}\\
\label{a19}&&\simeq\frac{1}{D^2\,\tilde{Y}^2}(1+\frac{\tilde{X}^2}{\tilde{Y}^2})(N^{\pi^+}+N^{\pi^-}),
\ee
where it is again adopted that distributions of hadrons $N^{\pi^+(\pi^-)}_{\plus(\minus)}$ are Poissonian,
so that $\delta(\Delta N^{\pi^+})=\delta(N^{\pi^+})=\sqrt{N^{\pi^+}_\minus+N^{\pi^+}_\plus}$,
$\delta(\Delta N^{\pi^-})=\delta(N^{\pi^-})=\sqrt{N^{\pi^-}_\minus+N^{\pi^-}_\plus}$ and one can neglect
${\rm cov}(\Delta N^{\pi^+},N^{\pi^-})$ and  ${\rm cov}(\Delta N^{\pi^-},N^{\pi^+})$.
By virtue of Eq. (\ref{virt}), 
\be
\left|\frac{\tilde X}{\tilde Y}\right|=\left|\frac{X_1-X_2}{X_1+X_2}\right|<<1,
\ee  
so that one can neglect\footnote{Let us recall (see footnotes \ref{foot13} and \ref{foot15}) 
that with the applied statistics $3\cdot10^6$ DIS events, 
the quantity $\tilde Y\equiv N^{\pi^+}-N^{\pi^-}$
essentially differs from zero even in the vicinity of the minimal value
$x_B=0.003$ accessible for measurement.
} 
$(\tilde X/\tilde Y)^2$ in Eq. (\ref{a19}). Thus, one again arrives at the approximate
formula (\ref{a16}) for the error on the difference asymmetries.

\end{document}